\documentclass[aps,prl,twocolumn,superscriptaddress,showpacs]{revtex4}
\usepackage{graphicx}
\usepackage{helvet}

\begin{document}

\title{Mott Transition in a Valence Bond Solid Insulator with a Triangular Lattice}

\author{Y. Shimizu}
\altaffiliation[Present address:]{Institute for Advanced Research, Nagoya University, Japan}
\affiliation{RIKEN, Wako, Saitama 351-0198, Japan.}

\author{H. Akimoto}
\affiliation{RIKEN, Wako, Saitama 351-0198, Japan.}

\author{H. Tsujii}
\altaffiliation[Present address:]{Department of Physics, Kanazawa University, Japan}
\affiliation{RIKEN, Wako, Saitama 351-0198, Japan.}

\author{A. Tajima}
\affiliation{RIKEN, Wako, Saitama 351-0198, Japan.}

\author{R. Kato}
\affiliation{CREST, RIKEN, Wako, Saitama 351-0198, Japan.}

\date{\today}

\begin{abstract}
	We have investigated the Mott transition in a quasi-two-dimensional Mott insulator EtMe$_3$P[Pd(dmit)$_2$]$_2$ with a spin-frustrated triangular lattice in hydrostatic pressure and magnetic field.
	In the pressure-temperature ($P$-$T$) phase diagram, a valence bond solid phase is found to neighbor the superconductor and metal phases at low temperatures. 
	The profile of the phase diagram is common to those of Mott insulators with antiferromagnetic order.
	In contrast to the antiferromagnetic Mott insulators, the resistivity in the metallic phase exhibits anomalous temperature dependence, $\rho = \rho_0 + A T^{2.5}$. 
	
\end{abstract}

\pacs{71.20.Rv, 71.30.+h, 74.25.Nf, 74.70.Kn}

\keywords{}

\maketitle
	
	The Mott transition has been one of the fundamental issues in condensed matter physics \cite{Mott1990}. 
	Much attention has been paid on the superconducting state near the Mott transition in transition metal compounds \cite{Imada1998} and in organic materials \cite{Kanoda2006}. 
	Mott insulators so far studied usually exhibit antiferromagnetic long-range order, and the relation between the existence of the ordered state and the appearance of superconducting states has been widely discussed \cite{Imada1998}. 
	When geometrical spin frustration is present, however, novel quantum states such as the resonating valence bond \cite{Anderson1973} and valence bond solid (VBS) states \cite{Read1989} can be realized. 
	Despite a considerable amount of theoretical work on the quantum antiferromagnets, only a few examples of the VBS state are known in two-dimensional (2D) systems, e.g., SrCu$_2$(BO$_3$)$_2$\cite{Kageyama1999} and BaCuSi$_2$O$_6$ \cite{Sebastian2006}, in which the metal-insulator transition has not been accessible. 
	Thus the Mott transition and superconductivity appearing from the quantum disordered state have been longed for the last few decades. 

	Here we show that an organic material is a good playground of the quantum spin physics on the frustrated lattice accessible to Mott physics. 
	An anion radical salt EtMe$_3$P[Pd(dmit)$_2$]$_2$ is the 2D VBS Mott insulator synthesized recently \cite{Kato2006}, where Et and Me denote C$_2$H$_5$ and CH$_3$ respectively, and Pd(dmit)$_2$ (dmit = 1,3-dithiole-2-thione-4,5-dithiolate, C$_3$S$_{5}^{2-}$) is an electron-acceptor molecule. 
	Its crystal structure consists of two layers: the Pd(dmit)$_2$ layer that involves in conduction and magnetism, and the insulating closed-shell EtMe$_3$P$^+$ layer. 
	In the conduction layer, pairs of Pd(dmit)$_2$ molecules form dimers arranged in a triangular lattice in terms of transfer integrals, $t$ and $t^\prime$ ($t^\prime/t$ = 1.05) [see Fig. 1 inset]. 
	The conduction band is half filling, consisting of an anitibonding combination of the highest-occupied molecular orbital of Pd(dmit)$_2$ \cite{Kato2004}. 
	The large onsite Coulomb interaction in the dimer, compared with the bandwidth, produces a Mott-Hubbard insulating state with a spin-1/2 at each dimer site. 
	The magnetic susceptibility behaves in accordance with the triangular-lattice antiferromagnetic Heisenberg model over a wide temperature range (60 K $< T <$ 300 K) with $J$ = 250 K \cite{Tamura2006}, indicating the presence of spin frustration and hence highly degenerate ground states. 
	The ground state is, however, settled into a spin-gapped phase by the lattice distortion via the spin-phonon coupling at 25 K, as shown in Fig. 1 inset \cite{Kato2006, Tamura2006}, while most of the Pd(dmit)$_2$ salts having anisotropic triangular lattice ($0.55<t^{\prime}/t<0.85$) show antiferromagnetic order \cite{Kato2004}.  
	Application of hydrostatic pressure, which increases the bandwidth, induces a superconducting transition at $T_{c}$ = 5 K \cite{Kato2006}. 
	However, the nature of the insulator and metal phases under pressure and the Mott transition between them are less understood.  
	In addition, how magnetic field affects the VBS on the frustrated triangular lattice and the Mott transition remains attractive problems of many-body quantum solids. 

	\begin{figure}
	\includegraphics[scale=0.4]{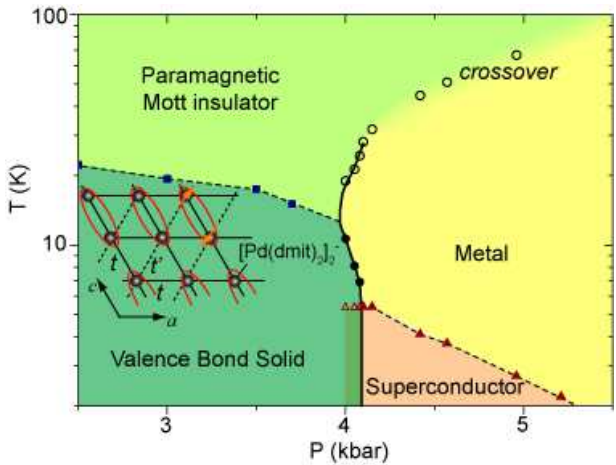}
	\caption{\label{Fig1} 
	$P$-$T$ phase diagram of EtMe$_3$P[Pd(dmit)$_2$]$_2$ obtained from resistivity measurements. 
	Open circle: metal-insulator transition/crossover temperatures; solid circle: insulator-metal transition temperatures (averages between the cooling and warming processes); square: VBS transitions; triangle: onset superconducting transitions. 
	The solid and doted lines denote the first-order and second-order transitions, respectively. 
	Inset: the schematic VBS state on the triangular lattice of [Pd(dmit)$_2$]$_2^{-}$. 
	}
	\end{figure}

	In this Letter, we report the pressure and magnetic field effects on the 2D VBS state of EtMe$_3$P[Pd(dmit)$_2$]$_2$ and demonstrate the $P$-$T$ phase diagram of the Mott transition, shown in Fig. 1. 
	We have found that the VBS state persists to the Mott boundary, where the superconducting state appears, without entering into the gapless spin liquid or antiferromagnetic ordered state. 
	However, the phase diagram of the 2D Mott transition is common to the Mott insulator with antiferromagnetic order.
	We discuss the unconventional electron scattering in the metallic state emerging from the VBS state based on the temperature dependence of resistivity.  

	Single crystals of EtMe$_3$P[Pd(dmit)$_2$]$_2$ were prepared by aerial oxidation of (EtMe$_3$P)$_2$[Pd(dmit)$_2$] in acetone containing acetic acid at 5 $^\circ$C. 
	A total of 6 samples with typical size of $0.8\times0.6\times0.1$ mm$^3$ were used for the resistivity measurements.
	A hydrostatic pressure was applied to the crystal in Daphne 7373 oil by using a BeCu pressure-cramp cell at room temperature. 
	The denoted pressure values are determined at room temperature by the resistance of the manganin wire in an accuracy of 0.01 kbar.
	The linearity between pressure and manganin resistance was calibrated by measuring a phase transition of Bi at 25.5 kbar. 
	Pressure is known to decrease by 1.5 kbar at low temperatures mainly due to oil solidification at 200-250 K but remain constant below 50 K \cite{Murata1997}. 
	We measured in-plane resistivity using a conventional four-probe method in direct current along the $a$ axis. 
	A magnetic field was applied to the $c$ axis, parallel to the conducting layer. 
	The temperature below 20 K and magnetic field were swept at rates of 0.1 K/min and 0.032 T/min, respectively. 
	The resistivity profiles in pressures are reproducible for different samples.

	\begin{figure}
	\includegraphics[scale=0.35]{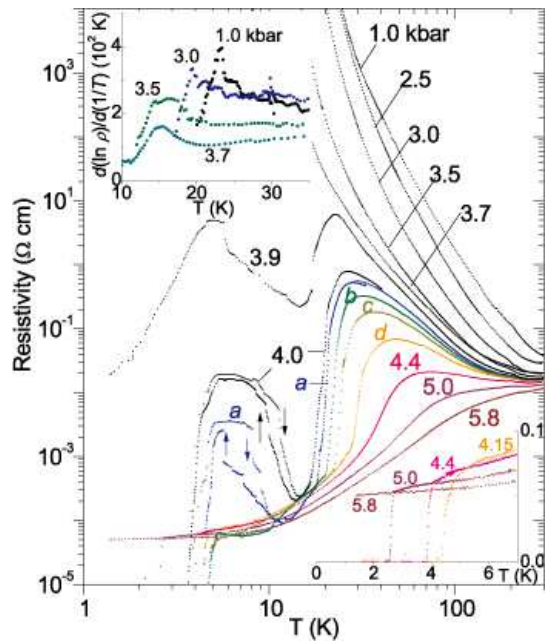}
	\caption{\label{Fig2} 
	Resistivity profile of EtMe$_3$P[Pd(dmit)$_2$]$_2$ in hydrostatic pressures ($a$ = 4.08, $b$ = 4.10, $c$ = 4.12, and $d$ = 4.15 kbar). 
	Inset: temperature dependence of the activation gap obtained as $\Delta$ = $d$(ln $\rho$)/$d(1/T)$.
	}
	\end{figure}

	Figure 2 shows the temperature dependence of resistivity at various pressures. 
	Below 3.7 kbar, the resistivity is semiconducting over the entire temperature range and does not exhibit thermal hysteresis or a resistance jump. 
	The activation gap given by $\Delta$ = $d$(ln $\rho$)/$d(1/T)$ shows a maximum as a function of temperature below 30 K (see inset of Fig. 2). 
	The temperature at a maximum $\Delta$, 23 K at 1.0 kbar, agrees with the observed VBS transition temperature at ambient pressure \cite{Tamura2006} and decreases to 15 K with increasing pressure up to 3.7 kbar, as also marked by the solid squares in Fig. 1. 
	The pressure dependence is consistent with the volume expansion of the lattice across the VBS transition at ambient pressure. 
	A key interest is whether the VBS state is fully suppressed by pressure until the Mott transition occurs.

	At 4.0 kbar, an abrupt resistivity drop of more than three orders of magnitude is observed in a narrow temperature range, 17 K $< T <$ 20 K, showing a first-order insulator-to-metal (IM) Mott transition in the bulk crystal. 
	Once the resistivity settles into metallic behavior, it increases again below 13 K. 
	A resistivity jump and a thermal hysteresis are manifestation of a first-order metal-to-insulator (MI) phase transition. 
	The reduction of resistivity below 5.3 K is attributed to the formation of percolated superconducting domains in the predominant insulating phase, while the similar resistivity behavior at 3.9 kbar arises from the formation of metallic current paths in the predominant insulating phase, which reflects the phase separation due to the first-order Mott transition. 
	Upon increasing pressure further, the reentrant MI transition is rapidly suppressed and disappears at 4.1 kbar.
	At the same time, the resistivity drop at 5.3 K becomes much sharper, manifesting the superconducting transition in the bulk. 
	The IM transition shifts to higher temperatures and becomes a crossover above 4.15 kbar and 30 K.
	 Above 5 kbar, the superconducting transition temperature decreases to below 1.5 K and the resistivity becomes metallic in all the measured temperature range. 

	These experimental results are summarized in Fig. 1, where the IM transition or crossover temperatures plotted by open circles are determined from the inflection point having the largest value of $d\rho/dT$ and the reentrant MI transition temperatures plotted by solid circles are defined from a position of a largest resistance jump. 
	At the triple point, where the insulator, superconductor and metal phases all meet, the two first-order transition lines ({\it i.e.} the insulator-metal and insulator-superconductor transitions) should be connected smoothly as shown in Fig. 1, since metal-superconductor transition is second order in zero field and $dP/dT$ must be 0 at $T=0$. 
	It is noteworthy that the positive $dT/dP$ slope of the first-order line at high temperatures turns negative at low temperatures, which strongly suggests the presence of spin order as seen in antiferromagnetic insulators \cite{McWhan1973, Lefebvre2000} because the Clausius-Clapeyron relation $dT/dP$ = $\Delta V/\Delta S$ ($<0$) requires that the insulator has smaller entropy than the metal, $\Delta S= S_{\rm ins} - S_{\rm metal}<0$, when the metal appears at high pressures, $\Delta V = V_{\rm ins} - V_{\rm metal}>0$. 

	\begin{figure}
	\includegraphics[scale=0.35]{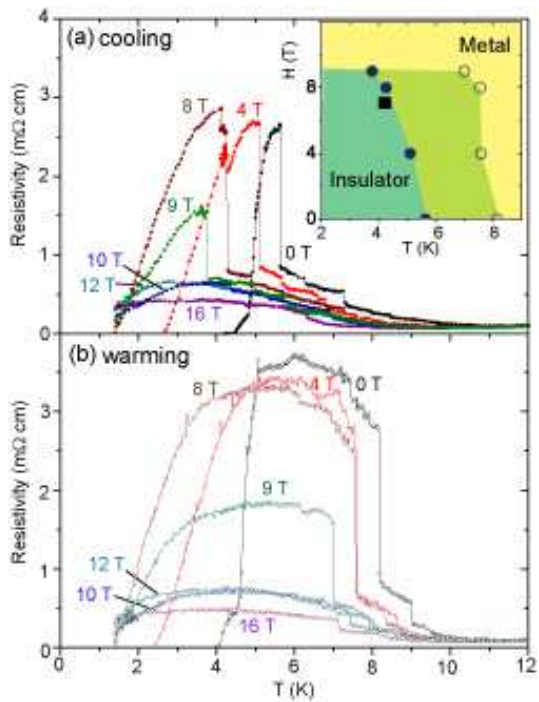}
	\caption{\label{Fig3} 
	Magnetic-field dependence of resistivity at 4.08 kbar for (a) cooling and (b) warming processes. 
	Inset: $H$-$T$ phase diagram, where solid and open circles are the MI transition temperatures for cooling and warming processes, and the square is the transition point of the field sweep measurement. 
	}
	\end{figure}

	We have investigated magnetic field effects on the Mott transition at 4.08 kbar, a critical pressure of the Mott boundary. 
	As shown in Fig. 3, in the absence of a magnetic field, the reentrant MI transition occurs principally at 5.6 K upon cooling, while the insulating state persists up to 8 K in the warming process. 
	The MI transition shifts to lower temperature with increasing magnetic field, and then disappears above 10 T for both the cooling and warming processes. 
	The magnetic field dependence of the resistivity at $T$ = 4.2 K is shown in Fig. 4. 
	For the field cooling at 16 T, the downward field sweep starts from the low-resistivity state at high field and then shows a resistivity jump to the high resistivity state at 7 T.
	Comparing with the $H$-$T$ diagram, the inset in Fig. 3, obtained from the temperature dependence of the resistivity at constant fields, the resistivity jump is attributable to the Mott transition, as marked by the square. 
	On the other hand, in the zero-field cooling case, the resistivity increases, with the magnetic field, to the high value dominated by the insulating state. 
	No indication of the Mott transition is seen up to 16 T, probably due to the large hysteresis of the transition. 
	The gradual increase of the resistivity from zero to high values is ascribed to the break down of the percolative superconducting clusters in the insulator.
	
	\begin{figure}
	\includegraphics[scale=0.4]{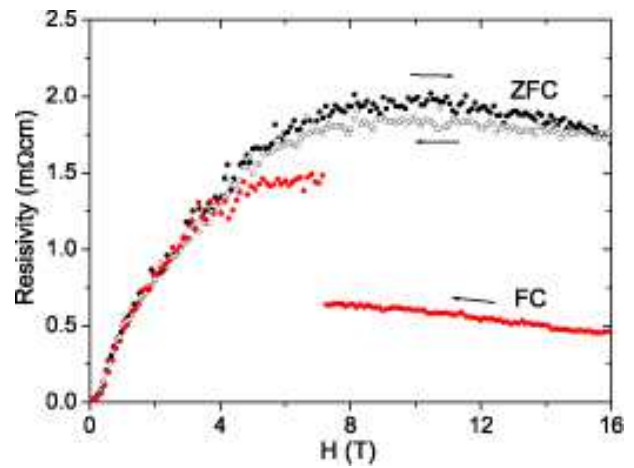}
	\caption{\label{Fig4} 
	Magnetoresistivity measured at 4.2 K and 4.08 kbar for zero-field-cooling (ZFC) and field-cooling (FC) at 16 T. 
	}
	\end{figure}

	From the slope of the phase boundary in the $H$-$T$ phase diagram, one can elucidate the magnetization difference between the insulating and metallic phases, $\Delta M =M_{\rm ins} - M_{\rm metal}$, utilizing a thermodynamic relation $dH/dT$ = $-\Delta S/\Delta M$. 
	Here $\Delta S<$ 0 as mentioned earlier and $dH/dT <$ 0 as shown in the inset of Fig. 3.
	These facts requires that $\Delta M < 0$, that is, the magnetization of the insulating phase must be smaller than that of the metallic phase.
	The result is distinct from ordinary Mott insulators having antiferromagnetic order, in which $\Delta M > 0$ induces the insulator-to-metal transition with increasing magnetic field \cite{Kagawa2004b, Vollhardt1984}. 
	Here $\Delta M < 0$ gives a strong restriction on the magnetism of the insulating state to be nonmagnetic. 
	The clear thermal hysteresis on the MI transition, seen in Fig. 2 and 3, is consistent with the simultaneous structural transition. 
	Thus, the main consequence from the observation of the field-induced transition is that the VBS state exists abut on the metal and superconducting phases. 
	The critical field of the IM transition, $H_{c}(0)$ = 9-10 T for the cooling process, yields the spin gap $\Delta = \textsl{g} \mu _{\rm B}H_{c}$ of, at least, 7-8 K. 
	To our knowledge, this is the first example of the pressure- and field-induced Mott transitions from the 2D VBS to the metal/superconductor, which accompany with the translational symmetry breaking of the lattice, similar to the Peierls transition from a metal to a band insulator. 
	Here the low-energy spin excitation in the Mott insulator is dominated by the local pair breaking of the valence bond, distinct from the antiferromagnetic long-range order.
	Nevertheless, the profile of the $P$-$T$ diagram is similar to that of the antiferromagnetic Mott insulator; namely, the superconducting phase emerges, independent of the magnetic ground state of Mott insulator, across the first-order Mott transition. 

	\begin{figure}
	\includegraphics[scale=0.4]{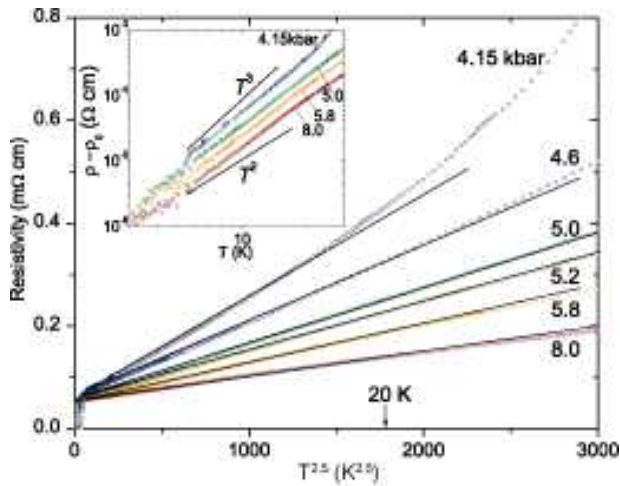}
	\caption{\label{Fig5} 
	Resistivity plotted as a function of $T^{2.5}$ in the metallic state. 
	The inset shows a log($\rho-\rho_0$) vs log$T$ plot at pressures below 30 K. 
	}
	\end{figure}

	The further intriguing issue is what the dominant interaction is in the metallic phase nearby the VBS phase, which could give an important insight into the superconducting pairing interaction. 
	Referring to the metals that neighbor on the antiferromagnetic Mott insulators, the low-temperature resistivity follows $\rho = \rho _0 + AT^{\varepsilon  } (\varepsilon = 2)$, as expected from the dominant electron-electron scattering in a standard Fermi liquid, where $\rho _0$ and $A$ are a residual resistivity and a constant that reflects the scattering rate between electrons \cite{Imada1998, Limelette2003}. 
	In the inset of Fig. 5, we plot the resistivity against temperature in double logarithmic scale after subtracting $\rho _0$. 
	We found that the resistivity follows a power law with 2 $< \varepsilon  <$ 3, for a wide range of pressures (4.15-8.0 kbar) at temperatures below 20-30 K. 
	Good linearity is found when the resistivity is plotted against $T^{2.5}$ as shown in the main panel of Fig. 5. 
	A deviation from $\varepsilon$  = 2 may indicate the marginal Fermi liquid or non Fermi liquid as discussed in heavy fermion systems \cite{Stewart2001} and cuprate superconductors \cite{Imada1998}. 
	The resistivity in these systems, however, commonly accompanies $\varepsilon$ smaller than 2. 
	The deviation to a larger value is unconventional and requires to consider contributions including the higher order electron scattering and the phonon effects. 
	Besides, it is noted that the resistivity and the $A$ value (0.6$\mu \Omega $cm/K$^2$, 4.15 kbar) are more than one order magnitude smaller than those of the other Mott systems (e.g., $A$=30$\mu \Omega $cm/K$^2$ for $\kappa $-(ET)$_2$Cu[N(CN)$_2$]Cl, 300 bar \cite{Limelette2003}), which indicates that the electron-electron scattering is not enhanced even near the Mott transition in EtMe$_3$P[Pd(dmit)$_2$]$_2$.
	The understanding of the nature of the metallic emerging from the VBS state needs further theoretical considerations.  
	The conceivable reasons for the reduced electron correlation, at the present stage, are the strong first-order nature of the Mott transition and the absence of the antiferromagnetic long-range order. 

	In conclusion, we established the Mott transition phase diagram of the VBS Mott insulator on the organic triangular lattice system. 
	On the Mott boundary, we demonstrated the presence of the VBS state, which is destabilized in a high magnetic field.  
	The similar phase diagram to the antiferromagnetic Mott systems suggests that the long-range magnetic order is not an essential factor for the emergence of superconductivity. 
	Furthermore, the suppressed electron-electron scattering in the metallic phase with the unconventional power law poses a question for the superconducting pairing interaction. 

	We thank N. Tajima, T. Itou, M. Tamura, Y. Ishii, M. Imada and S. Watanabe for useful discussions. 
	This work was supported by Grant-in-Aid for Scientific Research(No.16GS0219) from MEXT and JSPS.

\end{document}